\documentclass[conference, letterpaper]{IEEEtran}

\usepackage{mathtools, amssymb}
    \def\FF{\mathbb F}
    
    \def\leq{\leqslant}
    \def\AA{\mathcal A}
    \def\DD{\mathcal D}
    \def\OO{\mathcal O}
    \def\OZ{\mathcal O_Z}
    \newtheorem{theorem}{Theorem}
    \newtheorem{lemma}[theorem]{Lemma}

\usepackage{booktabs, tikz-cd}
    \usetikzlibrary{calc}

\usepackage[colorlinks, allcolors=blue!50!purple!90!black]{hyperref}

\usepackage{autonum}

\begin{document}
                                    
                                 \title
                      {Isolate and then Identify:
                 \\ Rethinking Adaptive Group Testing}
                                    
                                \author
       {\IEEEauthorblockN{Hsin-Po Wang and Venkatesan Guruswami}
                           \IEEEauthorblockA
       {Department of Electrical Engineering and Computer Science
            \\ Simons Institute for the Theory of Commputing
        \\ University of California, Berkeley, Berkeley, CA, USA
             \\ Emails: \{simple, venkatg\}@berkeley.edu}}
                                    
                               \maketitle

\begin{abstract}\boldmath
    Group testing (GT) is the art of identifying binary signals and the
    marketplace for exchanging new ideas for related fields such as
    unique-element counting, compressed sensing, traitor tracing, and
    geno-typing.  A GT scheme can be nonadaptive or adaptive; the latter
    is preferred when latency is ess of an issue.  To construct adaptive
    GT schemes, a popular strategy is to spend the majority of tests in
    the first few rounds to gain as much information as possible, and
    uses later rounds to refine details.  In this paper, we propose a
    transparent strategy called \emph{isolate and then identify} (I@I).
    In the first few rounds, I@I divides the population into teams
    until every team contains at most one sick person.  Then, in the
    last round, I@I identifies the sick person in each team.
    Performance-wise, I@I is the first GT scheme that achieves the
    optimal coefficient $1/$capacity$(Z)$ for the $k \log_2 (n/k)$ term
    in the number of tests when $Z$ is a generic channel corrupting the
    test outcomes.  I@I follows a modular methodology whereby the
    isolating part and the identification part can be optimized
    separately.
\end{abstract}

\section{Introduction}

    Group testing (GT) is the binary version of sparse signal
    regeneration whose goal is to find $k$ sick people among a
    population of $n$ using $m$ tests.  See Table~\ref{tab:models} for
    some applications of GT.  GT schemes can be benchmarked in many
    ways, the common ones being the following.

    \begin{itemize}
        \item To what extent are tests parallelizable, i.e., are they
            nonadaptive, two-round, or multi-round?
        \item How many tests do we need?  An entropic lower bound is
            $\Omega(k \log(n/k))$.  Is $\OO(k \log(n/k))$ or $\OO(k \log
            n)$ enough?
        \item Can the test results be noisy?  What if we let a binary
            symmetric channel (BSC) alter the test results?
        \item How long will decoding take?  Is $\OO(k \log n)$ time
            enough?
        \item How many mistakes will the decoder make?
    \end{itemize}

\begin{table}
    \caption{
        Problems related to group testing.
    }                                                 \label{tab:models}
    \centering
    \tabcolsep0pt
    \begin{tabular}{cccc}
        \toprule

        Problem
        & There are $n$ & $k$ of them & To solve, use $m$\\

        \midrule

        Disease control \cite{AlE22}
        & people & are sick & virus tests

        \\ Genotyping \cite{EGB10}
        & genes & cause cancer & gene tests

        \\ \scalebox{0.98}[1]{Wireless channel reservation} \cite{LuG08}
        & cellphones & want to talk & frequencies

        \\ \scalebox{0.98}[1]{Wireless direct transmission} \cite{LFP22}
        & messages & will be sent & frequencies

        \\ Heavy hitter \cite{ChN20}
        & items & are popular & words of storage
        
        \\ Compressed sensing \cite{ChN20}
        & signals & are active  & measurements

        \\ Differential privacy \cite{ACO22}
        & data points & are nonzero & words of space

        \\ DoS attack \cite{YIT10}
        & users & are spamming  & virtual servers

        \\ Traitor tracing \cite{CFN00}
        & users & resell keys & keys

        \\ Computer forensics \cite{GAT05}
        & files & will be modified & bits of storage

        \\ \scalebox{0.98}[1]{Property-preserving hashing \cite{Min22}}
        & properties & appears in a file & bits per file

        \\ Image compression \cite{HoL00}
        & coefficients & are nonzero & bits per layer

        \\ \bottomrule
    \end{tabular}
\end{table}

\begin{table*}
    \caption{
        Group testing works; $(\nu, \kappa) \coloneqq (\log_2 n, \log_2
        k)$.  $Z$ is a channel adding noise to test results; \\ $C(Z)$
        is its capacity; $\OZ$ is a constant depending on $Z$.  See
        \cite[Tables 2--7]{Gacha} for more details.
    }                                               \label{tab:capacity}
    \centering
    \begin{tabular}{ccccc}
        \toprule

        Name and reference
        & \#rounds & \#tests & $\DD$'s Complexity & Remark

        \\ \midrule

        Cheraghchi--Nakos \cite[Thm~20]{ChN20}
        & $1$ & $\OO(k \nu)$ & $\OO(k \nu)$ & $o(1)$ FPs\&FNs

        \\ Price--Scarlett \cite[Thm~1]{PrS20}
        & $1$ & $\OO(k \nu)$ & $\OO(k \nu)$ & $o(1)$ FPs\&FNs

        \\ Nonadaptive splitting \cite[Thm~1]{Bonsai}
        & $1$ & $\frac{1 + \varepsilon}{\ln2} k \nu$
        & $\OO(\varepsilon^{-2} k \nu)$ & $o(1)$ FPs\&FNs

        \\ Nonadaptive splitting \cite[Thm~2]{Bonsai}
        & $1$
        & $\frac{1 + \varepsilon}{\ln2} k \max(\nu - \kappa, \kappa)$
        & $\OO(\varepsilon^{-2} k^2 \nu)$ & $o(1)$ FPs\&FNs

        \\ SPIV \cite[Thm~1.2]{CGH21}
        & $1$ & $k \max(\nu - \kappa, \frac{\kappa}{\ln2})$
        & $n^{\OO(1)}$ & $o(1)$ FPs\&FNs

        \\ $2$-round SPIV \cite[Thm~1.3]{CGH21}
        & $2$ & $k (\nu - \kappa)$ & $n^{\OO(1)}$ & $o(1)$ FPs\&FNs

        \\ SPARC \cite[Theorem~1.1]{CHH24}
        & $1$ & $k (\nu - \kappa) / C(Z)$ & $n^{O(1)}$
        & $Z\colon \FF_2 \to \FF_2$; $o(k)$ FPs\&FNs

        \\ SPEX \cite[Theorem~1.3]{CHH24}
        & $1$ & $c_\text{ex} k (\nu - \kappa)$ & $n^{O(1)}$
        & $Z\colon \FF_2 \to \FF_2$; $o(1)$ FPs\&FNs
       
        \\ Bit-mixing coding \cite[Sec~V]{BCS21}
        & $1$ & $\OZ(k \nu)$ & $\OZ(k^2 \kappa \nu)$ & $o(1)$ FPs\&FNs
        
        \\ Noisy splitting \cite[Thm~4.1]{PST23}
        & $1$ & $\OZ(t k \nu)$
        & $\OZ((k (\nu - \kappa))^{1+\varepsilon})$
        & $k^{1-t\varepsilon}$ FPs\&FNs

        \\ Gacha GT \cite[Theorem~1]{Gacha}
        & $1$ & $\OZ(\sigma k \nu 2^{\OO(\tau)})$
        & $\OZ(\sigma k \nu^{\OO(1)} 2^{\OO(\tau)})$
        & $ k \exp(-\sigma \nu^{1-1/\tau})$ FPs\&FNs

        \\ SAFFRON \cite[Thm~7]{LCP19}
        & $1$ & $\OZ(k \nu)$ & $\OZ(k \nu)$ & $\varepsilon k$ FPs\&FNs

        \\ Nonadaptive GROTESQUE \cite[Cor~8]{CJB17}
        & $1$ & $\OZ(k \nu)$ & $\OZ(k \nu)$ & $\varepsilon k$ FPs\&FNs

        \\ Adaptive GROTESQUE \cite[Thm~1]{CJB17}
        & $\OO(\kappa)$ & $\OZ(k \nu)$ & $\OZ(k\nu)$ & $o(1)$ FPs\&FNs
        
        \\ Scarlett $2$-round \cite[Thm~1]{Sca19n}
        & $2$ & $k (\nu - \kappa) / C(Z) + \OZ(k\kappa)$
        & unspecified & $Z$ is BSC;  $o(1)$ FPs\&FNs


        \\ Scarlett $4$-round \cite[Thm~1]{Sca19a}
        & $4$ & $k (\nu - \kappa) / C(Z) + k \kappa /D_\text{KL}(p\|1-p)$
        & $\OZ(\varepsilon k^{2+\varepsilon})$
        & $Z$ is BSC$(p)$; $o(1)$ FPs\&FNs

        \\ Teo--Scarlett splitting \cite{TeS22}
        & $m$ & $k (\nu - \kappa) / C(Z) + \OZ(k\kappa)$
        & presumably low & BSC; zero-error

        \\ \cmidrule(l{3em}r{3em}){1-5}

        \textbf{Isolate and then Identify} [Thm~\ref{thm:main}]
        & $\OO(\kappa)$
        & $k (\nu - \kappa) / C(Z) + \OO(k \kappa / (1 - 2p)^2)$
        & $\OO(k \nu^2 / C(Z)^2)$ & $Z$ is BSC$(p)$; $o(k)$ FPs\&FNs \\

        \textbf{I@I generic channel} [Thm~\ref{thm:generic}]
        & $\OO(\kappa)$
        & $k (\nu - \kappa) / C(Z) + \OZ(k \kappa)$
        & $\OO(k \nu^2 / C(Z)^2)$ & $Z$ is generic; $o(k)$ FPs\&FNs \\

        \bottomrule
    \end{tabular}
\end{table*}

    Combining together all benchmarks,
    a natural challenge is to construct a perfect
    GT scheme that is fully parallelizable, uses $\OO(k \log n)$ tests,
    resists noises, decodes in $\OO(k \log n)$ time, and makes mistake
    with probability $\to 0$.  However, this problem remains open as
    recent works, some listed in Table~\ref{tab:capacity}, are only able
    to find GT schemes that give up some criteria to leave room for the
    others.  For instance, nonadaptive splitting \cite{ChN20, PrS20,
    Bonsai} requires noiseless tests.  Bit-mixing coding \cite{BCS21},
    noisy splitting \cite{PST23}, and SPARC and SPEX \cite{CHH24} relax
    decoding complexity.  Teo--Scarlett's splitting \cite{TeS22}
    is not parallelizable at all.
    
    In this work, we give up some parallelizability to prioritize the
    other criteria.  Commonly, adaptive GT schemes follow the design
    pattern below (including SPIV \cite{CGH21}, GROTESQUE \cite{CJB17},
    and Scarlett's \cite{Sca19n, Sca19a}).
    \begin{itemize}
        \item In the first few rounds, apply $k \log_2 (n/k)$ tests
            to obtain $\hat x \in \{0, 1\}^n$, a coarse-grained
            approximation to the actual indicator $x \in \{0, 1\}^n$ of
            sick people.
        \item In later rounds, apply $\varepsilon k \log n$ tests to
            refine $\hat x$.
    \end{itemize}
    In contrast, this paper advocates for a different design pattern.
    \begin{itemize}
        \item In the first few rounds, we divide the population into
            teams until each team contains at most one sick person.
        \item In the last round, we identify the sick person in each
            team. 
    \end{itemize}
    We call this pattern \emph{isolate and then identify} (I@I).  It is
    the main contribution of our paper and is formally stated below.

    \begin{theorem}[main]                               \label{thm:main}
        Let $Z$ be a BSC with crossover probability $p$ that models the
        noisy test outputs; let $C(Z) \coloneqq 1 + p \log_2(p) + (1 -
        p) \log_2(1 - p)$ be its capacity.  Suppose $n/k \to \infty$.
        To find $k$ sick people among $n$, I@I uses $(1 + o(1)) k
        \log_2(n/k) / C(Z) + \OO(k \log(k) / (1 - 2p)^2)$ tests in
        $\OO(\log k)$ rounds.  I@I's decoder uses $\OO(k \log(n)^2 /
        C(Z)^2)$ time and produces $o(k)$ false positives (FPs) and
        false negatives (FNs) on average.
    \end{theorem}

    The rest of this paper is dedicated to proving
    Theorem~\ref{thm:main} and discussing implications.
    Section~\ref{sec:state} states the problem formally.
    Section~\ref{sec:isolate} describes the isolating part of I@I.
    Section~\ref{sec:identify} details the identifying part of I@I
    and concludes the proof of Theorem~\ref{thm:main}.
    Section~\ref{sec:imply} discusses how to generalize
    Theorem~\ref{thm:main} to generic channels and other open-ended
    problems.

\section{Formal Problem Statement}                     \label{sec:state}

    Let there be $n$ people.  Let $x \in \{0, 1\}^{n\times1}$ be an
    indicator vector where $x_j = 1$ if the $j$th person is sick and
    $x_j = 0$ if healthy, $j = 1, \dotsc, n$.  Let there be $k$ sick
    people and hence $k$ ones in $x$.  We assume that $k$ is known in
    advance.\footnote{ If not, there are GT-like schemes that count the
    number of sick people \cite{Bsh24}.} Any size-$k$ subset of $[n]$
    has the same probability to become the set of sick people.

    An $r$-round adaptive GT scheme is a sequence of $r$ algorithms
    $\AA^{(1)}, \AA^{(2)}, \dotsc, \AA^{(r)}$.  Here, $\AA^{(1)}$ takes
    no input and return a binary matrix $A^{(1)} \in \{0, 1\}
    ^{m^{(1)}\times n}$.  We then use $A^{(1)}$ to perform $m^{(1)}$
    tests
    \[
        y^{(1)} \coloneqq \min(A^{(1)}x, 1)
        \in \{0, 1\}^{m^{(1)}\times1},
    \]
    where $A^{(1)}x$ is matrix multiplication and the minimum function
    replaces anything greater than $1$ by $1$.  Let $Z$ be a binary
    symmetric channel (BSC) with crossover probability $p$.  Let
    $z^{(1)} \coloneqq Z(y^{(1)}) \in \{0, 1\}^{m^{(1)}\times1}$ be the
    noisy vector where each entry of $y^{(1)}$ is post-processed by an
    iid copy of Z.

    For $q = 2, \dotsc r$, inductively, $\AA^{(q)}$ takes $z^{(1)},
    \dotsc, z^{(q-1)}$ as inputs and outputs $A^{(q)} \in \{0,
    1\}^{m^{(q)}\times n}$, which is used to perform $m^{(q)}$ tests
    \[
        y^{(q)} \coloneqq \min(A^{(q)}x, 1)
        \in \{0, 1\}^{m^{(q)}\times1},
    \]
    but the algorithms and the decoder sees the noisy version $z^{(q)}
    \coloneqq Z(y^{(q)})$.

    The decoder $\DD$ is an algorithm that takes $z^{(1)}, \dotsc,
    z^{(r)}$ as inputs and outputs $\hat x \in \{0, 1\}^{n\times1}$ as a
    guess of $x$.  The performance of $\DD$ is measured by $\|\hat x -
    x\|_1$, the number of mistakes, averaged over the randomness of $x$,
    $\AA^{(1)}, \dotsc, \AA^{(r)}$, and $\DD$.  The cost of this GT
    scheme is measured by the number of rounds $r$, the total number of
    tests $m \coloneqq m^{(1)} + \dotsb m^{(r)}$ (as a random variable),
    and the complexities\footnote{Very often the complexity of $\DD$
    dominates those of $\AA^{(1)}, \dotsc, \AA^{(r)}$ so the complexity
    of the decoder alone is a good benchmark.} of $\AA^{(1)}, \dotsc,
    \AA^{(r)}$ and $\DD$.
    
\section{The Isolating Part of I@I}                 \label{sec:isolate}

    We explain the isolating part of I@I in this section.  The goal of
    this part is to divide the population into teams such that each team
    contains zero or one sick person.

\subsection{The first round of I@I}

    To begin, divide the population into $k$ teams, each team containing
    $n/k$ people.  Since the sick people are randomly distributed, the
    number of sick people in each team is a binomial random variable
    with $n/k$ trials and success probability $k/n$.  Since $n$, $k$,
    and $n/k$ are supposedly large, readers might as well assume that
    the number of sick people in each team is a Poisson random variable
    with mean $1$.

    For each team, we plan to spend $s$ tests, for some $s$ that will be
    determined shortly, to learn if the number of sick people is $0$,
    $1$, or $2+$.  To do so, let each of the $s$ tests randomly includes
    half of the team.  That is, each row of the testing matrix $A^{(1)}$
    will have $n/2k$ ones that are placed randomly.  The teams can be
    categorized into one of the three types.
    \begin{itemize}
        \item Empty team: If a team contains no sick people, no tests
            will be positive.
        \item Exact team: If a team contains exactly one sick person,
            about $s/2$ tests will be positive.
        \item Twoplus team: If a team contains two or more sick people,
            about $3s/4$ tests (or even more) will be positive.
    \end{itemize}
    But recall that test results are noisy---they are flipped by $Z$
    with crossover probability $p$.  Therefore, the three team types
    correspond to the following three expectations.
    \begin{itemize}
        \item For an empty team, about $p$ tests will look positive.
        \item For an exact team, about $1/2$ tests will look positive.
        \item For a twoplus team, at least $(1-p)3/4 + p/4 = 3/4 - p/2$
            tests will look positive.
    \end{itemize}

    The problem we are facing now is that a test result is a Bernoulli
    random variable, of which we take $s$ samples, and we want to guess
    if the mean is $p$, $1/2$, or $3/4 - p/2$.  This is a classical
    hypothesis testing problem.  By the standard Hoeffding inequality,
    the probability that we misclassify the mean is
    $2^{-\Theta((1/2-p)^2s)}$.  To control the misclassification
    probability to be $< 1/k^3$, it is clear that
    \[ s \coloneqq \Theta(\log(k) / (1 - 2p)^2) \]
    is necessary and sufficient.

    To recap, this is what the first round of I@I does: spend $k s =
    \Theta(k \log(k) / (1 - 2p)^2)$ tests to learn how many sick people
    ($0$, $1$, or $2+$?) are in each of the $k$ teams.  Since the
    misclassification probability, $1/k^3$, is rather small, we proceed
    with the assumption that the classification of teams into $0$, $1$,
    and $2+$ is correct.

\subsection{The subsequent rounds}

    By the Poisson approximation, there are about $k/e$ empty teams,
    about $k/e$ exact teams, and about $k(1 - 2/e)$ twoplus teams.  What
    we want to do in the second round of I@I is to re-divide the
    twoplus teams to further isolate sick people.

    Merge all twoplus teams to form a big team.  Let $k^{(1)}$ denote
    the number of sick people in this big team.  This is $k$ minus the
    number of exact teams, which can be approximated as $k^{(1)} \approx
    (1 - 1/e) k$.  We then re-divide this big team into $k^{(1)}$ teams,
    randomly.  This is very similar to the situation we faced during the
    first round: Back then, we had $k$ sick people and $k$ random teams;
    now, we have $k^{(1)} \approx (1 - 1/e) k$ people and $k^{(1)}
    \approx (1 - 1/e) k$ random teams.

    Now for the second round of I@I, we spend $s = \Theta(\log(k) / (1
    - 2p)^2)$ tests on each team to learn if it contains $0$, $1$, or
    $2+$ sick people.  Apply the same hypothesis testing argument and
    infer that we miscount with probability $1/k^3$ per team.  Because
    the miscount probability is small, we proceed with the assumption
    that the classification into $0$, $1$, and $2+$ is correct.

    For the third round of I@I, we merge the twoplus teams obtained in
    the second round and let $k^{(2)}$ be the number of sick people
    therein.  It can be calculated that $k^{(2)}$ is $k^{(1)}$ minus the
    number of exact teams in the second round, which is about $k^{(1)} /
    e$.  Therefore, $k^{(2)} \approx (1 - 1/e) k^{(1)} \approx (1 -
    1/e)^2 k$.  We re-divide that into $k^{(2)}$ teams and spend $s =
    \Theta(\log(k) / (1 - 2p)^2)$ tests on each team to perform the
    classification into $0$, $1$, and $2+$.

\begin{figure}
    \centering
    \begin{tikzpicture}[x=0.8cm]
        \draw (0, 0) node (root) [below, draw] {$n$ people, $k$ sick};
        \foreach \j in {-5, ..., 5}{
            \draw (\j-0.5, -1) rectangle node{
            \ifnum \j < -1
                $0$
            \else\ifnum \j < 2
                $1$
            \else
                $2+$
            \fi\fi
            } (\j+0.5, -1.5);
        }
        \foreach \j in {-1, ..., 5}{
            \draw (\j-1, -2) rectangle node{
            \ifnum \j < 0
                $0$
            \else\ifnum \j < 3
                $1$
            \else
                $2+$
            \fi\fi
            } (\j+0, -2.5);
        }
        \foreach \j in {0, ..., 5}{
            \draw (\j-1.5, -3) rectangle node{
            \ifnum \j < 1
                $0$
            \else\ifnum \j < 5
                $1$
            \else
                $2+$
            \fi\fi
            } (\j-0.5, -3.5);
        }
        \foreach \j in {3, ..., 5}{
            \draw (\j-2, -4) rectangle node{
            \ifnum \j < 3
                $0$
            \else\ifnum \j < 6
                $1$
            \else
                $2+$
            \fi\fi
            } (\j-1, -4.5);
        }
        \draw [dotted, line width=1pt]
            (root.south west) -- (-5.5, -1)
            (root.south east) -- (5.5, -1)
            (1.5, -1.5) -- (-2, -2)
            (5.5, -1.5) -- (5, -2)
            (2, -2.5) -- (-1.5, -3)
            (5, -2.5) -- (4.5, -3)
            (3.5, -3.5) -- (1, -4)
            (4.5, -3.5) -- (4, -4)
        ;
    \end{tikzpicture}
    \caption{
        The isolating part of I@I.  A row is a round.  Empty teams are
        discarded.  Exact teams are kept for the last round.  Twoplus
        teams are re-divided.  In this case, $(k, k^{(1)}, k^{(2)},
        k^{(3)}, k^{(4)}) = (13, 10, 7, 3, 0)$
    }                                                    \label{fig:cut}
\end{figure}
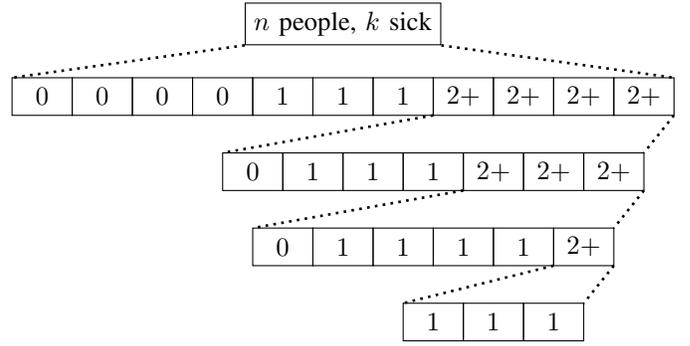

    We repeat the same process for sufficiently many rounds, each round
    merging and re-dividing the twoplus teams into $k^{(q)}$ teams,
    where $k^{(q)}$ is the number of sick people after $q$ rounds.  The
    number of sick people still belonging to twoplus teams will decay
    exponentially as $q$ increases and eventually vanish.  See
    Figure~\ref{fig:cut} for an illustration.  We dedicate the rest of
    this section to correctness analysis, including why $\Theta(\log k)$
    rounds is sufficient.  Readers interested in the GT specification
    more than the correctness proofs may jump to the next section.

\subsection{Analysis of the isolating part}

    To analyze the isolating part of I@I, it is essential to study the
    random variables $k, k^{(1)}, k^{(2)}, \dotsc$ as they control the
    number of rounds and the total number of tests. To study that, let
    $r_j$ be the index of the round in which the $j$th sick person is no
    longer in a twoplus team, $j = 1, \dotsc, k$.  We have the following
    claim. 
    \begin{lemma}
        The number of rounds it takes to isolate all sick
        people and the total number of tests are
        \[
            \max_{j\in[k]} r_j
            \qquad\text{and}\qquad
            s \sum_{j\in[k]} r_j,                  \label{for:maxr-sumr}
        \]
        respectively.
    \end{lemma}

    \begin{IEEEproof}
        After $r \coloneqq \max_{j\in[k]} r_j$ rounds, no sick people
        are still in twoplus teams.  So we finish the isolation part at
        the $r$th round.

        For the number of tests, note that the $q$th round spends
        $sk^{(q)}$ tests.  In total $s(k + k^{(1)} + k^{(2)} + \dotsb +
        k^{(r)})$ tests are spent.  Think of $k + k^{(1)} + k^{(2)} +
        \dotsb + k^{(r)}$ as the Lebesgue sum and think of
        $\sum_{j\in[k]} r_j$ as the Riemann sum: They are both counting
        pairs $(j, q)$ where the $j$th sick person is still in some
        twoplus team after $q$ rounds.  Therefore, the total number of
        tests used is $s(k + k^{(1)} + k^{(2)} + \dotsb + k^{(r)}) = s
        \sum_{j\in[k]} r_j$.
    \end{IEEEproof}

    \smallskip

    To control \eqref{for:maxr-sumr} using $n$ and $k$, it suffices to
    study the tail probability of $r_j$ and the concentration of the sum
    of $r_j$.

    The tail probability of $r_j$ can be controlled as follows.  A sick
    person must sit in one team and it is up to the other sick people if
    they accidentally join the same team.  Using the Poisson
    approximation, we know that the probability that the other sick
    people accidentally join the same team is about $1 - 1/e$.  A
    conservative upper bound of that is $2/3$.  The probability that a
    sick person is still in a twoplus team after $r$ rounds is
    $(2/3)^r$.  Hence $r_j$ is at most a geometric random variable with
    success probability $1/3$.  Give it $\Theta(\log k)$ rounds, we can
    reduce that probability to $< 1/k^3$.  Since there are $k$ sick
    people, the probability that any $r_j$ is $> \Theta(\log k)$ is $<
    1/k^2$.  We therefor assume that the isolating part will end in
    $\Theta(\log k)$ rounds.  If there are still twoplus teams after
    $\Theta(\log k)$ rounds, we simply declare failure.

    The concentration of $\sum_j r_j$ follows a similar logic.  By the
    nature of geometric distribution, the sum of $k$ geometric random
    variables is $> 4k$ if and only if $4k$ Bernoulli trials succeed $<
    k$ times.  But since the success probability is $1/3$, it is
    exponentially unlikely that this happens.  We infer that we should
    declare failure when we see $\sum_j r_j > 4k$, and the failure
    probability would be $< e^{-\Omega(k)}$.  In any case, the isolating
    part of I@I spends $\OO(k) s = \OO(k \log(k) / (1 - 2p)^2)$ tests,
    failure or not.

\section{The Identifying Part of I@I}              \label{sec:identify}

    In Section~\ref{sec:isolate}, we invested $\OO(k \log(k) / (1 -
    2p)^2)$ tests performed in $\OO(\log k)$ rounds to isolate the sick
    people into exact teams.  In this section, we explain the
    identifying part of I@I.  Note that this part shares ideas with
    GROTESQUE \cite{CJB17}, SAFFRON \cite{LCP19}, and Gacha
    \cite{Gacha}.

\subsection{Use code to identify the sick person}

    For any exact team, prepare $\ell$ tests and a good error correcting
    code over the binary field with code dimension $\log_2(\text{team
    size}) \leq \log_2(n/k)$ and block length $\ell$; we will determine
    $\ell$ very shortly.  A choice is Barg and Zémor's codes
    \cite{BaZ02}, which achieve capacity, assume positive error
    exponents, and decode in linear time.
    
    We now establish a bijection that maps each person to a
    codeword and let this person participate in tests at places where
    the codeword has one.  In other words, each column of the testing
    matrix $A^{(r)}$ will be a codeword of the code (plus padding zeros
    at places allocated to other teams).

    Given the construction above, the test results will be a codeword
    that corresponds to the sick person; the noisy test results will be
    a corrupted codeword caused by the BSC $Z$.  Because Barg and
    Zémor's codes achieve capacity, we can take
    \[ \ell \coloneqq (1 + \varepsilon) \log_2(n/k) / C(Z) \]
    and enjoy a per-block error probability of $\varepsilon$ for some
    $\varepsilon > 0$ that converges to $0$ as $n/k \to \infty$.  With
    this, we can give a formal proof of the main theorem.

\subsection{Proof of Theorem~\ref{thm:main}}

    \begin{IEEEproof}
        The overall design is as follows: Use the adaptive procedure
        introduced in Section~\ref{sec:isolate} to divide the sick
        people into teams such that each team contains either zero (so
        this team will be discarded) or one (so this team will be passed
        to the identifying part) sick person.  For the identifying part,
        use Barg and Zémor's codes to identify the sick person in each
        team, as explained in the last subsection.

        The number of rounds is upper bounded by design: In the
        isolating part, we declare failure when the number of rounds is
        $> \Theta(\log k)$.  The total number of tests is also upper
        bounded by design: In the isolation part, we declare failure
        when the number of tests reaches $\OO(k)s = \Theta(k \log(k) /
        (1 - 2p)^2)$.  In the identifying part, we only allocate $k \ell
        = (1 + o(1)) k \log_2(n/k) / C(Z)$ tests.  Therefore, the total
        number of tests is always $\leq (1 + o(1)) k \log_2(n/k) / C(Z)
        + \Theta(k \log(k) / (1 - 2p)^2)$.

        We next analyze the sources of FNs and FPs.

        When we declare failure, we stop any testing plan and report $k$
        random people to be sick.  This will produce at most $k$ FNs and
        $k$ FPs.  But since we declare failure with probability
        $\OO(1/k^2)$, this only contribute $\OO(1/k)$ FNs and FPs on
        average.

        The second source of FPs and FNs in the isolating part is that
        we assume that classification of teams into $0$, $1$, and $2+$
        is always correct.  Since each classification is wrong with
        probability $1/k^3$ and there are $\OO(k \log k)$
        classifications, the probability of at least one
        misclassification is $\OO(\log(k) / k^2)$.  Since it is not
        possible to produce more than $k$ FPs and $k$ FNs, on average we
        produce $\OO(\log(k) / k)$ FPs and FNs.

        The last possible source of FNs and FPs is the identifying part.
        Because each exact team produces $2 \varepsilon$ FPs and FNs on
        average, where $\varepsilon \to 0$, all $k$ exact teams together
        produces $o(k)$ FPs and FNs on average.

        As for the decoding complexity, we assert without going over
        details that the complexity of the the isolating part is $\OO(k
        \log k)$.  For the identifying part, Barg and Zémor's codes has
        decoding complexity linear in the block length for any fixed
        code rate.  But because we want a series of capacity-achieving
        codes, we need to relax the linear complexity to, say, quadratic
        complexity.  We conclude the theorem with the complexity of the
        identifying part being $\OO(k \ell^2) = \OO(k
        \log(n/k)^2/C(Z)^2)$.
    \end{IEEEproof}

\subsection{Customize the identifying part}

    In this subsection, we discuss several ways to customize
    Theorem~\ref{thm:main}.

    The first customization we want to discuss is that one might want to
    relax the assumption $n/k \to \infty$ to $n \to \infty$.  In other
    words, the only restriction on $k$ would be $1 \leq k\leq n$.  To do
    so, we increase the block length to $\ell \coloneqq (1 +
    \varepsilon) \* \log_2(n) / C(Z)$.  This means that there will be
    $(1 + \varepsilon) \* k \* \log_2(k) / C(Z)$ extra tests.

    A further customization is to increase $\ell$ to $c \log_2(n) /
    C(Z)$ for some reasonable multiplier $c > 0$.  This way, the code
    rate of Barg--Zémor is always less than $C(Z)/c$ and we will observe
    a positive error exponent.  That is, I@I will produce fewer FNs and
    FPs; in fact, it will produce $o(1)$ FNs and FPs on
    average\footnote{ With Markov's inequality, this means that the
    probability of having any FN or FP at all is $o(1)$.}
    if\footnote{This if is always the case when $c \to \infty$.} the
    error exponent at rate $C(Z)/c$ is greater than $1/c$.

    Given the observation made in the last paragraph, another way to
    reduce the number of FNs and FPs is borrowed from earlier adaptive
    GT works \cite{Sca19n, Sca19a}: use a few more rounds to
    double-check whether the people identified by the identifying part
    are really sick, and re-identify if not.  More precisely, with
    $\OO(k \log(k) / (1 - 2p)^2)$ tests in the first extra round we can
    identify the $\varepsilon k$ misidentified healthy people.  And with
    $c \varepsilon k \log_2(n) / C(Z)$ tests in the second extra round
    we can reduce the number of FNs and FPs from $o(k)$ to $o(1)$.

\section{Generalization and Open Problems}             \label{sec:imply}

    In the last subsection, we see that the straightforward approach of
    I@I makes it very easy to customize.  In this section, we go one
    step further and consider generalizing the main theorem to arbitrary
    channels.

\subsection{Generic channels}

    By a generic channel we mean a triple $Z = (\Sigma, \mu_0, \mu_1)$
    where $\Sigma$ is the output alphabet, $\mu_0$ is the law of the
    output when the input is $0$, and $\mu_1$ is the law of the output
    when the input is $1$.  Under the new channel model, the test result
    at the $q$th round is still $y^{(q)} \coloneqq \min(A^{(q)}x, 1) \in
    \{0, 1\}^{m^{(q)}\times1},$ but the observation \[ z^{(q)} \coloneqq
    Z(y^{(q)}) \in \Sigma^{m^{(q)}\times1} \] is no longer binary.  An
    example of $Z$ is an additive Gaussian noise channel; its $\Sigma$
    is the real line and its $\mu_0$ and $\mu_1$ are normal
    distributions $\mathcal{N}(1, \sigma^2)$ and $\mathcal{N}(-1,
    \sigma^2)$ respectively.

    We now state the generalized main theorem.

    \begin{theorem}[generic channel]                 \label{thm:generic}
        Let $C(Z)$ be the capacity of $Z$; let $D(Z)$ be some constant
        depending on $Z$ (which we will specify in the proof).  Suppose
        $n/k \to \infty$.  To find $k$ sick people among $n$, I@I uses
        $(1 + o(1)) k \log_2(n/k) / C(Z) + \OO(k \log(k) / D(Z))$ tests
        in $\OO(\log k)$ rounds.  I@I spends $\OO(k \log(n)^2) / C(Z)^2
        + \OO(k \log(k)) / D(Z)$ time to decode and produces $o(k)$ FPs
        and FNs on average.
    \end{theorem}

    \begin{IEEEproof}
        The proof is almost identical to that of the main theorem: we
        spend the first few rounds to isolate the sick people and use
        the last round to identify the sick person in each exact team.
        In the following, we explain the differences only.

        Let's discuss the identification part first as it is rather
        straightforward: we replace Barg and Zémor's codes by polar
        codes.  Polar codes achieve capacity over any binary-in many-out
        channel with decoding complexity $\OO(\ell \log \ell)$
        \cite{HoY13}, where the block length in use is $\ell = (1 +
        \varepsilon) \log_2(n/k) / C(Z)$.  We relax the complexity bound
        to $\OO(\ell^2)$ and conclude that $k$ exact teams take $\OO(k
        \ell^2) = \OO(k \log(n)^2 / C(Z)^2)$ time to decode.  The rest
        of the identification part of the proof remains unchanged.

        Let's discuss the isolation part now.  For isolation, we want to
        determine $s$, the number of tests used to classify each team so
        that the misclassification probability is at most $1/k^3$.  This
        problem boils down to determining the exponent $D(Z)$ that
        controls the error probability $2^{-D(Z)s}$ of using hypothesis
        testing to distinguish distributions $\mu_0$ and $\frac{\mu_0 +
        \mu_1} 2$ and $\frac{\mu_0 + 3\mu_1}4$, where $s$ is the number
        of samples.  We can take $D(Z)$ to be the minimum of the four KL
        divergences
        \begin{gather}
            D_\text{KL}(\mu_0 \| \tfrac{\mu_0+\mu_1}2), \;
            D_\text{KL}(\tfrac{\mu_0+\mu_1}2 \| \mu_0), \\
            D_\text{KL}(\tfrac{\mu_0+\mu_1}2 \| \tfrac{\mu_0+3\mu_1}4),
            \text{ and }
            D_\text{KL}(\tfrac{\mu_0+3\mu_1}4 \| \tfrac{\mu_0+\mu_1}2).
        \end{gather}
        We then take $s \coloneqq \Theta(\log(k) / D(Z))$.  The rest of
        the isolation part of the proof remains unchanged.
    \end{IEEEproof}

\subsection{Open-ended problems}

    We close this paper by raising a couple of problems that, when
    answered, lead to better versions of I@I.

    The first problem is to find the optimal value of $D(Z)$.  More
    precisely, if we let every test in the isolation part includes
    $fn/k$ random people for some free parameter $f$, then the
    distributions we want to distinguish become
    \[
        \mu_0
        \text{ versus }
        (1 - f) \mu_0 + f \mu_1
        \text{ versus }
        (1 - f)^2 \mu_0 + (2f - f^2) \mu_1
    \]
    It is unclear if there are explicit expressions or bounds for $f$
    and $D(Z)$.

    The second problem is to balance the number of rounds and tests in
    the isolation part.  More precisely, as the sequence $k^{(1)},
    k^{(2)}, k^{(3)}\dotsc$ decay to zero, we can re-divide the twoplus
    teams into $2k^{(1)}, 3k^{(2)}, 4k^{(3)}, \dotsc$ teams (or maybe
    $2k^{(1)}, 4k^{(2)}, 8k^{(3)}, \dotsc$ teams) to accelerate
    isolation at the cost of slightly more tests.  A primary calculation
    shows that we might be able to reduce the number of rounds
    from$\OO(\log k)$ to $\OO(\log(\log k))$.

    The third problem is even more open-ended.  We have demonstrated
    that it is straightforward to apply a code to find the sick person
    when we know there is only one.  We can even do so with the optimal
    number of tests $\log_2(\text{team size}) / C(Z)$.  But, as long as
    there are more sick people, even if there are only two, the
    difficulty surges.  Existing approaches include list-decoding
    \cite{DoW22} (which only works in a limited range of parameters) or
    enumerating over some large set \cite{BCS21} (but then the error
    probability is bound to the set's size).  We seek new ideas on this
    general direction.

\section{Conclusion}

    In this paper, we studied a design pattern for adaptive GT.  During
    the design, we modularly decompose the task of GT into isolation and
    identification, hence the name I@I.  This is analogous in spirit to
    certain real-root--finding algorithms that break the real line into
    intervals, each containing at most one root, and apply the
    Newton--Raphson method to find the only root in each interval.

    I@I is a low-test, low-complexity GT scheme.  But its biggest
    advantage is perhaps that of being transparent and easy to
    customize.  We demonstrate this point by considering arbitrary
    channels and obtain the optimal coefficient $1/C(Z)$ in front of the
    $k \log(n/k)$ term (which is the main term if $k$ is fixed and $n
    \to \infty$).  Several open problems remain.  One concerns the
    optimal value $D(Z)$ can take for each channel, another concerns
    whether the isolation part can be done in fewer rounds, and the
    third problem concerns identifying two sick people in a team.

\section{Acknowledgment}

    Research supported in part by
    NSF grant CCF-2210823 and a Simons Investigator Award.

\IEEEtriggeratref{15}
\bibliographystyle{IEEEtran}
\bibliography{Measure1Cut1-30}

\end{document}